\documentstyle[11pt]{article} 
\def\thefootnote{\fnsymbol{footnote}}
\def\L{{\cal L}}

\def\Tr{{\,\mbox{Tr}}}

\def\del{{\partial}}
\def\bea{\begin{eqnarray}}
\def\eea{\end{eqnarray}}
\def\be{\begin{eqnarray}}
\def\ee{\end{eqnarray}}
\def\beq{\begin{equation}}
\def\eeq{\end{equation}}
\def\ben{\begin{eqnarray}}
\def\een{\end{eqnarray}}

\def\noin{\noindent}

\def\Tr{{\,\rm Tr}}
\def\del{\partial}
\def\L{{\cal L}}

\def\I{{\cal I}}

\def\Msol{M_{\mbox{\small sol}}}

\def\Lsu{{\L_{\mbox{\tiny SU(2)}}}}

\def\vsl{{v \!\!\! /}}
\def\d{{\rm d}}

\begin{document}

\begin{titlepage}
\setcounter{footnote}{1}
\begin{flushright}
{{\large \bf SNUTP-92/78}}
\end{flushright}

\vskip 2.5cm

\begin{center}
{\Large \bf Soliton Structure of Heavy Baryons} \\
\vskip 1.6cm
{Dong-Pil MIN$^{\mbox{\small a}}$, Yongseok OH$^{\mbox{\small a}}$,
Byung-Yoon PARK$^{\mbox{\small b}}$}\\
and \\ Mannque RHO$^{\mbox{\small c}}$

\vskip 1.5cm

\end{center}

\noindent \hskip 0.5cm{\it $^{\mbox{\small a}}$
Department of Physics and Center for Theoretical Physics, \\
\mbox{\hskip 0.7cm} Seoul National University, Seoul 151--742, Korea} \\
\ \mbox{\hskip 0.5cm}{\it  $^{\mbox{\small b}}$ Department of Physics,
Chungnam National University,
  Taejeon 302-764, Korea} \\
\ \mbox{\hskip 0.5cm}{\it $^{\mbox{\small c}}$ Service de Physique
 Th\'{e}orique, C.E. Saclay, 91191 Gif-sur-Yvette, France.} \\

\begin{center}
\vskip 1.5cm
{\bf ABSTRACT}
\vskip 0.3cm

\begin{quotation}
\noin
Heavy-quark baryons are described as a bound heavy-meson-soliton
system in a Lagrangian that combines chiral symmetry and
heavy-quark symmetry.  We introduce a ``Wess-Zumino type" term and
show that it dominates the binding
of a heavy meson to a soliton. The connection between
this model and the Callan-Klebanov model is established to
$O(N_c^{-1} \cdot m_\Phi^0)$ where $m_\Phi$ is the mass of the
heavy-meson (isospin) doublet $\Phi$ or $\Phi^*$.
\end{quotation}
\end{center}

\end{titlepage}
\setcounter{footnote}{0}
\def\thefootnote{\#\arabic{footnote}}

\newpage

It has recently been shown\,\cite{RRS,OMRS}
that the skyrmion description\,\cite{CK} of heavy baryons
as one or more heavy pseudoscalar mesons in isospin doublet ``wrapped" by
an $SU(2)$ soliton works surprisingly
well not only for strange hyperons but also for
charmed as well as bottom hyperons provided one takes empirical values
for the decay constants and masses of the pseudoscalar mesons. The results
for spectra and magnetic moments were found to be remarkably close to
the results of quark models which are expected to fare better heavier
the quark involved. Analogy to the induced gauge potentials
that describe the excitations of diatomic molecules has led to the
suggestion\,\cite{mr} that the hyperfine (and fine) splitting
for baryons $(Qqq)$ where
$Q$ represents a heavy quark and $q$ a light quark of flavor up and down
takes the form
\be
\Delta E_{hf}\sim \frac{1}{2\I} (\vec{J}_l + c \vec{J}_Q )^2,
\label{MOL}
\ee
where ${\cal I}$ is the moment of inertia of the soliton,
$\vec{J}_l$ stands for the angular momentum lodged in the
light-quark (soliton) system and $\vec{J}_Q$ carried by heavy meson
denoted generically $\Phi$ and that in the limit that the heavy-quark mass
$m_Q$ or equivalently the heavy-meson mass containing $Q$ denoted $m_\Phi$
goes to infinity, the hyperfine coefficient $c$ goes to zero and hence the
heavy-quark spin $J_Q$ decouples\footnote{The analogy to the diatomic
molecule is seen when the interatomic distance $R$ goes to infinity at
which limit the angular momentum of the induced gauge field decouples.}.
Such a limiting behavior would be consistent with the heavy-quark symmetry
of Isgur and Wise\,\cite{IW}. Unfortunately up to date, we have been unable
to derive (\ref{MOL}) from the point of view of induced gauge structure
or to show from the Callan-Klebanov (CK) formulation that the coefficient $c$
indeed has the right asymptotic property.

The purpose of this paper is to examine the heavy-quark limit of
the skyrmion description by taking the heavy-meson limit on the effective
Lagrangian used in Refs.\cite{RRS,OMRS} and comparing with a Lagrangian
recently constructed by Wise\,\cite{Wise} and Yan {\it et al.}\,\cite{Yan}
to satisfy both the chiral symmetry of light quarks and the Isgur-Wise symmetry
of heavy quarks. We show in particular that in the heavy-quark limit,
$c$ vanishes. Our work overlaps closely with -- and in part is
stimulated by -- the recent work of Jenkins, Manohar and Wise (JMW)\,\cite{JMW}
and Guralnik {\it et al.}\,\cite{GLM}
on the structure of heavy baryons at the fine- and hyperfine-structure level.
There is however a distinct difference from Refs.\cite{Wise}$\sim$\cite{GLM}
in that we start with chiral symmetry in the CK framework and approach
the heavy-quark limit from below. In the CK approach, the soliton contributes
to the heavy baryon mass a term of $O(N_c^1)$,
the binding energy of the soliton and meson contributes at
$O(N_c^0)$ and while the fine and
hyperfine splitting occurs at $O(N_c^{-1})$,
arising from the collective rotation of the soliton {\it and} the bound meson,
it is {\it formally} at $O(m_\Phi^0)$. Thus
the standard $N_c$--counting is still valid in the infinite mass limit.
In Ref.\cite{GLM}, the hyperfine splitting occurs at $O(m_\Phi^{-1}\cdot
N_c^{-1})$. We will argue shortly that there is no disagreement on this
since in our approach, there is a hidden $m_\Phi^{-1}$ dependence in the
hyperfine coefficient $c$ of (\ref{MOL}).

We start with a chiral Lagrangian that contains vector mesons together
with the chiral field. One could make a rather general discussion
using a hidden gauge symmetric (HGS) Lagrangian of
Bando {\it et al.}\,\cite{Bando,Fujiwara}, but for our purpose,
it suffices to slightly
modify and study the model of Ref.\cite{SMNR}
which has proven to be phenomenologically successful. In the notation
suitable to our purpose, the Lagrangian can be written as
the sum of the $SU(2)$ Skyrme Lagrangian, $\Lsu$, the HGS Lagrangian
without (with) the $\omega$ meson coupling, $\L_\Phi^{\mbox{\tiny HGS}}$
$(\L_\omega^{\mbox{\tiny HGS}})$, and the ``anomalous parity"
 Lagrangian, ${\L}_{an}$;
\be
\L^{\mbox{\tiny HGS}} &=& \Lsu + \L_\Phi^{\mbox{\tiny HGS}}
               + \L_\omega^{\mbox{\tiny HGS}} + \L_{an},
\nonumber \\
\Lsu &=& \frac{F_\pi^2}{16}
\Tr \left[ \del_\mu \Sigma \del^\mu \Sigma^\dagger
\right] + \frac{1}{32e^2} \Tr \left[ \Sigma^\dagger \del_\mu \Sigma ,
\Sigma^\dagger \del_\nu \Sigma \right]^2 ,   \nonumber \\
\L_\Phi^{\mbox{\tiny HGS}} &=&
\left(D_\mu \Phi \right)^\dagger \left( D^\mu \Phi \right)
- m_\Phi^2 \Phi^\dagger \Phi \nonumber \\
&& - \frac{1}{2} \Phi^{*\dagger}_{\mu\nu} \Phi^{*\mu\nu} +
m_{\Phi^*}^2 \left[ \Phi^{*\dagger}_\mu + \frac{2i}{F_\pi g_{\Phi^*}}
\Phi^\dagger A_\mu \right] \left[ \Phi^{*\mu} + \frac{2i}{F_\pi g_{\Phi^*}}
A^\mu \Phi \right]       \nonumber \\
\L_\omega^{\mbox{\tiny HGS}} &=& -\frac{iN_c}{2F_\pi^2} B_\mu
\left[ \left(  \Phi^\dagger D^\mu \Phi
       - (D^\mu \Phi)^\dagger \Phi \right)
 - \left( \Phi_\nu^{*\dagger} D^\mu \Phi^{*\nu}
- (D^\mu \Phi^*_\nu)^\dagger \Phi^{*\nu} \right) \right] \nonumber \\
{\L}_{an} &=&-  \frac{iN_c}{F_\pi^2} B_\mu
\left(  \Phi^\dagger D^\mu \Phi
       - (D^\mu \Phi)^\dagger \Phi \right)+\delta {\L}_{an}
\label{Lag:HGS}
\ee
where
\be
D_\mu &=& \del_\mu + V_\mu, \hskip 1cm \Sigma = \xi \cdot \xi, \nonumber \\
\left( \begin{array}{c} V_\mu \\ A_\mu \end{array} \right) &=& \frac{1}{2}
\left( \xi^\dagger \del_\mu \xi \pm \xi \del_\mu \xi^\dagger \right),
\nonumber \\
B_\mu &=& \frac{1}{24\pi^2} \epsilon_{\mu\nu\alpha\beta} \Tr \left\{
\Sigma^\dagger \del^\nu \Sigma \Sigma^\dagger \del^\alpha \Sigma
\Sigma^\dagger
\del^\beta \Sigma \right\}, \nonumber \\
\Phi^*_{\mu\nu} &=& \del_\mu \Phi^*_\nu - \del_\nu \Phi^*_\mu
+ V_\mu \Phi_\nu^* - V_\nu \Phi_\mu^*,
\ee
with $\epsilon_{0123}=+1$. The Lagrangian ${\L}_{an}$ which
contains, in addition
to the usual Wess-Zumino term \cite{witten}, intrinsic-parity-odd
four-derivative terms involving vector fields requires some explanation
to which we will return below.
Here, $\Sigma$ is the $SU(2)$ chiral field,
$\Phi$ and $\Phi^*_\mu$  are, respectively, the pseudoscalar and vector
meson doublets of the form $Q\bar{q}$, $F_\pi$ represents
the pion decay constant and
$g_{\Phi^*}$ is the $\Phi^*$ ``gauge" coupling to matter fields.
The Skyrme parameter $e$ will be specified later.
For instance, if we take the kaons to be heavy mesons,
$\Phi^\dagger = (K^-, \overline{K}^0)$,
$\Phi_\mu^{*\dagger} = (K_\mu^{*-}, \overline{K}_\mu^{*0})$.
This Lagrangian is obtained from that of Ref.\cite{SMNR} by
integrating out the $\omega$ and $\rho$ meson fields and then
taking the limit
$m_\Phi= m_{\Phi^*} \to \infty$, neglecting the terms that
vanish as $m_\Phi^{-1}$ and $m_{\Phi^*}^{-1}$ or faster. For the purpose of
comparing with the Isgur-Wise symmetric limit, it is necessary to keep the
vector mesons explicitly instead of integrating them out as we did in
Ref.\cite{OMRS}. The reason for this will become clear later on.

We need to explain a bit what ${\L}_{an}$ is in the context of the heavy-meson
limit that we are interested in. The first term is what one obtains from the
topological Wess-Zumino term written down by Witten \cite{witten} when expanded
\`{a} la Callan-Klebanov. This is intrinsically tied to anomalies in
effective theory. Later, as the heavy quark mass increases, this term
will disappear.\footnote{M.A. Nowak and I. Zahed,
private communication. This point is discussed further in \cite{lnrz}.}
However the second term, which is intrinsic-parity odd
as the Wess-Zumino term is and involves the vectors $P^*$'s, needs not
vanish in the heavy-quark limit. We expect them to modify the constants
of the main term responsible for the binding of the mesons $\Phi$ and
$\Phi^*$ to a soliton. As we know from the work of \cite{Bando,Fujiwara},
they are fixed at low energy by low-energy theorems. They can be fixed
in the heavy-quark limit by the heavy-quark symmetry.

To see what remains of the Lagrangian (\ref{Lag:HGS}) in the
heavy-quark limit, we make the meson-field redefinition as in Ref.\cite{Yan},
(taking $m = m_\Phi = m_{\Phi^*}$)
\be
\Phi^{*\dagger}_\mu &=& e^{-im v \cdot x} P_\mu^* / \sqrt{m}, \nonumber \\
\Phi^\dagger &=& e^{-im v \cdot x} P / \sqrt{m},
\label{field}
\ee
so that the fields $P_\mu^*$ and $P$ are independent of the
meson mass and obtain
\be
\L_\Phi^{\mbox{\tiny HGS}} &=&
- i P v \cdot \stackrel{\leftrightarrow}{D} P^\dagger
+i P^*_\mu v \cdot \stackrel{\leftrightarrow}{D} P^{*\mu\dagger}
+ i \sqrt{2} \left( P A^\mu P^{*\dagger}_\mu
+ P^*_\mu A^\mu P^\dagger \right),
\label{HGS-lag} \\
\L_\omega^{\mbox{\tiny HGS}} &=&
\frac{N_c}{F_\pi^2} B_\mu
\left( P v^\mu P^\dagger - P_\nu^{*} v^\mu P^{*\nu\dagger}
\right) \label{LWZT} \\
{\L}_{an} &=& 2\frac{N_c}{F_\pi^2} B_\mu P v^\mu P^\dagger
+ \delta {\L}_{an}\label{LWZ}
\ee
where
\be
(D_\mu P)^\dagger &=& ( \del_\mu + V_\mu ) P^\dagger, \nonumber \\
A \stackrel{\leftrightarrow}{D} B^\dagger
&=& A (DB)^\dagger - (DA) B^\dagger.
\ee
We have not written out the term $\delta {\L}_{an}$ since while the
coefficients are known phenomenologically in the light-quark sector,
they are not known in the regime we are concerned with. We expect that
it will include terms of the form
\be
\frac{iN_c}{F_\pi \pi^2} \epsilon^{\mu\nu\alpha\beta} v_\mu
\left(a P A_\nu A_\alpha P_\beta^{*\dagger} -
b P_\beta^* A_\alpha A_\nu P^\dagger
\right)
\ee
with $a$ and $b$ unknown constants \footnote{For light-quark baryons,
low-energy
theorems fix it to $a=b=1/3$. For heavy-quark baryons, the Isgur-Wise symmetry
requires that $a=b$ but the overall constant is not known. See later.}.
Note that since in SMNR\,\cite{SMNR}
we started with an apparently $SU(3)$ symmetric Lagrangian
(apart from the meson mass term)
with the flavor $Q$ put on the same footing as the light quarks,
Eq. (\ref{LWZT}) results from the $\omega$-meson
coupling terms and hence the constant $N_c/F_\pi^2$ is fixed.
In the heavy-quark limit, the heavy pseudoscalar
decouples from the Wess-Zumino term,
so the first term of Eq.(\ref{LWZ}) which comes from the primordial
Wess-Zumino term will disappear. However, the second term will
remain to modify effectively
the coefficient of Eq.(\ref{LWZT}) which came
from the $\omega$-meson coupling with the heavy mesons $P$ and $P^*$.
That such a term must be present can be seen by bosonizing light and
heavy quarks from QCD \cite{NRZ}.

If the $P$ and $P^*$ are degenerate, there are additional terms
in an HGS Lagrangian that can contribute at the same order in the
normal as well as anomalous parts of the Lagrangian.
One such term comes from the intrinsic-parity
odd term accompanying the Wess-Zumino term in the
HGS Lagrangian\,\cite{Fujiwara} that survives
in the heavy-meson mass limit
\be
c_4 \L_{(4)} = - c_4 2i g_{\Phi^*}^2
\epsilon^{\mu\nu\alpha\beta} v_\mu P_\nu^{*} A_\alpha P^{*\dagger}_\beta,
\label{four-d}
\ee
with $c_4$ the coefficient of $\L_{(4)}$\,\cite{Bando}. This is effectively
a four-derivative term that belongs to the same intrinsic-parity class
as $\delta {\L}_{an}$ discussed above. The coefficient $c_4$ is fixed
in the light-quark sector to $c_4=iN_c/16\pi^2$ from the decay
$\omega\rightarrow \rho \pi$\cite{Fujiwara}.
We will determine later its value in the
heavy-quark sector.

Our Lagrangian (\ref{HGS-lag})--(\ref{four-d}) can now be
compared with the one used by JMW\,\cite{JMW}
\be
\L_\Phi^{\mbox{\tiny JMW}} &=& -i \Tr \overline{H}_a v^\mu \del_\mu H_a
 + i \Tr \overline{H}_a H_b v^\mu \left(V_\mu \right)_{ba} \nonumber \\
& & + ig \Tr \overline{H}_a H_b \gamma^\mu \gamma_5
\left( A_\mu \right)_{ba} + \cdots \ \ ,  \label{LH}
\ee
with the heavy meson field $H_a$ (where $a$ labels the light-quark flavor)
defined as
\be
H = \frac{ ( 1 + \vsl )}{2} \left[ {P^*}_{\mu} \gamma^\mu -
P \gamma^5 \right]. \label{Ha}
\ee
Substituting Eq.(\ref{Ha}) into (\ref{LH}) and taking the trace over the gamma
matrices, we have\,\cite{Yan}
\be
\L_\Phi^{\mbox{\tiny JMW}}
&=& - i P v \cdot \stackrel{\leftrightarrow}{D} P^\dagger
+ i P^{*}_\mu v \cdot \stackrel{\leftrightarrow}{D} P^{*\mu\dagger}
\nonumber \\
&& + 2ig \left\{ P^{*}_\mu A^\mu P^\dagger + P A^\mu P_\mu^{*\dagger} \right\}
+ 2g \epsilon^{\lambda\mu\nu\kappa} v_\lambda P^{*}_\mu A_\nu
P^{*\dagger}_\kappa.
\label{Lag:Yan}
\ee
Ignoring for the moment the term (\ref{LWZT}) which we will
take up shortly, we see that the SMNR Lagrangian (\ref{HGS-lag}) with
(\ref{four-d}) is identical to the JMW Lagrangian (\ref{Lag:Yan}) if we
identify $g = 1 / \sqrt{2}$\footnote{This is rather close to the
quark-model prediction $g\simeq 0.75$\,\cite{JMW} and also to the
phenomenological value $g\approx 0.6$ extracted from the CLEO
collaboration data \cite{cho}.}
and $c_4 g_{\Phi^*}^2=ig$.
JMW did not take into account the term (\ref{LWZT})
since it involves higher derivatives. This procedure is perhaps justified
in the meson sector since $B_\mu$ involves three derivatives and hence
is suppressed by the factor
$\frac{p^3}{\Lambda^3}$ where $\Lambda$ is the chiral scale of order
$m_\rho$ and $p$ is of order $m_\pi$. However in the nontrivial topological
sector $B_\mu$ is the baryon current and is of $O(1)$. Furthermore
in the Callan-Klebanov scheme, the terms (\ref{LWZT})
and (\ref{LWZ}) play the key role in
binding the heavy pseudoscalar doublet to the soliton. We therefore
propose to add to the JMW Lagrangian (\ref{LH}) the
$\L_\omega^{\mbox{\tiny HGS}}$,
Eq.(\ref{LWZT}), preserving chiral
and Isgur-Wise symmetries
\be
\L^{\mbox{\tiny HGS}}_\omega &=&  \alpha B_\mu j^\mu, \\
j^\mu &=& \Tr \left( \overline H v^\mu H \right).\label{LWZP}
\ee
with $\alpha$ that as noted above cannot be determined by symmetries
alone.
Such a term arises in an approximate bosonization of QCD, through the coupling
of $H$ to the $\omega$ meson\,\cite{NRZ} so there is no reason to ignore it.
Although this has nothing to do, at least
in the heavy-quark limit, with the {\it bona-fide} Wess-Zumino term,
we will refer, {\it for convenience}, to it as ``Wess-Zumino
type" term to suggest that in the limit that $m_Q$  goes to zero,
it would have the same form as -- and perhaps be linked to --
the {\it topological} Wess-Zumino term.
In (\ref{LWZP}), $j^\mu$ is the $U(1)$ current of the
Lagrangian $\L_\Phi^{\mbox{\tiny JMW}}$ corresponding to the heavy-quark
flavor which is conserved in our case.
Although as mentioned above, the coupling constant $\alpha$ cannot be
determined
by chiral and Isgur-Wise symmetries alone, we will analyze the structure
of heavy baryons in units of $-N_c / 2F_\pi^2$, {\it i.e.}, the coefficient of
${\L}_\omega^{\mbox{\tiny HGS}}$ in Eq.(\ref{LWZT}).
We will normalize the meson field as
\be
\int \d^3 r j^0 = -2  \int \d^3 r
\left( P P^\dagger + P_i^{*} P^{*\dagger}_i \right)
= -1 \label{normalization}
\ee
and work in the rest frame of the heavy meson, $v_\mu = (1,0,0,0)$.
Note that $P_0^* = 0$ due to $v \cdot P^* = 0$.

The Lagrangian correct to order $O(m_\Phi^0 \cdot N_c^0)$ is given by
\be
L_B &=&- \Msol-m_\Phi + \int \d^3 r( {\cal L}_P+ \L_W ) ,   \nonumber \\
-{\cal L}_P &=&  2gi \left\{ P^{*i} A^i P^\dagger
+ P A^i P^{*i\dagger}
- i \epsilon^{0ijk} P^{*i} A^j P^{*k\dagger} \right\} \nonumber \\
- \L_W &=&  2 \alpha B_0 \left( P P^\dagger + P^{*}_i P^{*\dagger}_i
\right).
\label{Ham}
\ee
One can readily see that ${\cal L}_P$ and ${\cal L}_W$ are invariant with
respect to the global
rotation $S\in SU(2)_V$ in the light flavor space ({\it i.e.}, the isospin
space) provided that the fields transform
\be
P(x) &=&  \phi (x) S^\dagger,  \nonumber \\
P_i^* (x) &=&  \phi_i^* (x)  S^\dagger, \nonumber \\
\xi (x) &=& S \xi_0 (\vec x) S^\dagger,
\label{trans}
\ee
with $x = (t,\vec x)$ and $\xi_0 (\vec x) = \exp ( i \vec \tau \cdot \hat r
F(r) / 2)$ in the hedgehog configuration. We should stress that
in order for $L_B$ to be invariant, the transformation
(\ref{trans}) is {\it required}. The standard procedure for collective
quantization is to
elevate $S$ to a dynamical variable by endowing it with the time dependence
$S(t) = a_0 (t) + \vec a (t) \cdot \vec \tau$. Note that as defined, the
fields $\phi (x)$ and $\phi_i^* (x)$ are fields living in the rotating
frame\footnote{It should be noted that the quantization followed in
Refs.\cite{JMW,GLM} differs from our procedure.
In Refs.\cite{JMW,GLM}, the heavy
meson fields are {\it not} rotated, so are not defined in the soliton rotating
frame. Instead they are defined in their rest frame. There is of course
nothing wrong in their procedure and it explains why they need not resort
to the isospin-spin transmutation. The heavy mesons there are not
{\it behaving} like heavy quarks as in our approach.
It would be interesting, however, to understand why the sign
change of the coupling constant $g$ or of the radial shape function
$F(r)$ affects the binding in Refs.\cite{JMW,GLM}.}.

{}From the equations of motion for $\phi (x)$ and $\phi^* (x)$ gotten from
the Lagrangian valid at $O(m_\Phi^0 \cdot N_c^0)$, one can readily arrive
at the feature that the probabilities of the pseudoscalar and vector mesons
are peaked at the center of the soliton
\be
| \phi (x) |^2  &\propto& \delta ^3 (\vec{x}), \nonumber\\
|\phi^*_i (x)|^2  &\propto& \delta ^3 (\vec{x}).\nonumber
\ee
That this must be so can be understood as follows. In the large $N_c$
and large $m_\Phi$ (or equivalently large $m_\Phi^*$) limit, the soliton
and the meson will be on top of each other and hence when
the soliton is fixed at the origin, the wavefunction of both $P$ and $P^*$
in the inertial frame of the rotating soliton must be of the delta function
type. Given these solutions, it is now a simple matter to calculate the
energy shift
coming from $-({\cal L}_P + \L_W)$ of (\ref{Ham})
\be
E_I &=&- \int ({\cal L}_P + \L_W) \d^3 r \nonumber \\
&=& - \frac{1}{2\pi^2} \alpha  \left\{ F'(0) \right\}^3 .
\label{energy}
\ee
An important point to note here is
that the contribution of $\L_P$ is {\it zero}\,\cite{future}.
In fact, this $\L_P$ term is proportional to the scalar product of
the isospin vector and the spin vector of the light degrees of freedom
of $H$ field\,\cite{GLM}, {\it i.e.}, $\vec I_H \cdot \vec S_{\ell H}$.
Contrary to Ref.\cite{GLM}, however, in our formalism,
to $O(N_c^0)$ the
light degrees of freedom in $H$ field do not possess isospin and
spin quantum numbers. The field $H$ gets its quantum numbers only after
collective rotation. Therefore in the rotating frame their
``expectation values" must be zero. This is guaranteed in our calculation
at $O(N_c^0)$ by the exact cancellation among the three terms in ${\cal L}_P$.
This shows the difference between
the quantization procedure adopted here and that
adopted by JMW\,\cite{JMW,GLM}.

As suggested above, we take $g = 1/\sqrt{2}
\simeq 0.7$, $F^\prime (0)\approx
-0.89 $ GeV from the literature and the experimental value of
$F_\pi = 186$ MeV and $e=4.75$, with which
we find the $\alpha$ value in the b-quark sector should be
\footnote{The $\alpha$ so obtained is numerically not very different
from the value if we take $\alpha = - N_c / (2 F_\Phi^2)$
with $F_D=1.84 F_\pi$ and $F_B=1.67 F_\pi$. These values are consistent
with those employed in Ref.\cite{OMRS}.}
\be
\alpha \approx - \frac{1}{2.8} \left( \frac{N_c}{2 F_\pi^2} \right)
\ee
to reproduce $M_{\Lambda_b} - M_{\mbox{\tiny N}} = 4.65$ GeV which is the
predicted value of the quark model. This corresponds to the fine splitting of
\be
E_I\approx -0.55 {\mbox{ GeV}}.
\ee

Next we consider the effects of $O(m_\Phi^0 \cdot N_c^{-1})$ term in the
Lagrangian. For this we define
\be
S^\dagger \dot S = i \vec \tau
\cdot \vec \Omega
\ee
and write the corresponding Lagrangian to $O(m_\Phi^0 \cdot N_c^{-1})$
\be
L_{(-1)} = \int \d^3 r \L_{(-1)} = 2 \I \Omega^2 - 2 \vec \Omega \cdot \vec Q,
\label{canonical}
\ee
where
\be
\vec Q &=& - \int \d^3 r \left( \phi \vec n (\vec r) \phi^\dagger +
\phi_i^{*} \vec n (\vec r) \phi_i^{*\dagger} \right), \nonumber \\
\vec n &=& \frac{1}{2} \left( \xi_0^\dagger \vec \tau \xi_0 + \xi_0 \vec \tau
\xi_0^\dagger \right) \nonumber \\
&=& \cos F(r) \vec \tau - (\cos F(r) - 1 ) \hat r ( \vec \tau \cdot \hat r ),
\ee
and $\I$ is the moment of inertia of the $SU(2)$ soliton determined from
the properties of the $N$ and $\Delta$.
As suggested by JMW\,\cite{JMW},
because of the $\delta$-function structure of
the meson wavefunctions and a parity-flip at the origin,
it is more convenient to transform the heavy-meson fields to
\be
\phi &\to& \phi' =  \phi \, \xi_0 , \nonumber \\
\phi^*_\mu &\to& \phi_\mu^{*'} = \phi_\mu^* \, \xi_0, \nonumber \\
\vec n &\to& \xi_0\,  \vec n \, \xi^\dagger_0.
\label{fieldredef}
\ee
Note that the binding energy is not affected by this transformation.
With the primed fields, $\vec Q$ is of the form
\be
\vec Q &=& - \frac{1}{2} \int \d^3 r
\left\{ \phi' \left( \Sigma^\dagger \vec \tau
\Sigma + \vec \tau \right) \phi'^\dagger + \phi^{*'}_i
\left( \Sigma^\dagger \vec \tau
\Sigma + \vec \tau \right) \phi^{*'\dagger}_i \right\}.
\label{Qprime}
\ee
Now since in the soliton rotating frame, the
``isospin" of the meson is transmuted to spin, we can identify
\be
\vec Q = c \vec J_Q,
\ee
namely, proportional to
the angular momentum lodged in the meson which is $1/2$.
Canonical quantization of (\ref{canonical}) leads to an
$O(m_\Phi^0 \cdot N_c^{-1})$ splitting \footnote{Modulo a hidden $m_\Phi^{-1}$
dependence in $c$ explained below.} in energy given
by (\ref{MOL})\,\cite{SMNR},
\be
\Delta E_{hf} &=& 2 \I \Omega^2   \nonumber \\
&=& \frac{1}{2\I} \left\{ c J(J+1) + (1-c)J_\ell(J_\ell+1)
+ c(c-1) J_Q (J_Q+1) \label{hfsplit}
\right\},
\ee
where $\vec J_\ell$ is the spin lodged in the rotor as discussed
in Ref.\cite{SMNR}.  The total spin $\vec J$ of the system is
\be
\vec{J}=\vec J_\ell + \vec J_Q.
\ee
By an explicit calculation\,\cite{future}, we find that with (\ref{Qprime})
\be
c=0 ,
\ee
where we used the normalization of (\ref{normalization}) which is
invariant under the transformation (\ref{fieldredef}). The way we arrive at
this result is quite intriguing and highly nontrivial.
The first term of (\ref{Qprime})
coming from the $P$ mesons gets cancelled exactly by the second
coming from the $P^*$ mesons. If the $P$ and $P^*$ were not degenerate
the cancellation would not occur. This suggests the following scenario.
For not too large $m_\Phi$, say, $m_K$, $c$ can be substantial,
of $O(1)$, since the $K^*$ is rather high-lying compared with the
$K$. As $m_\Phi$ becomes large, the $P^*$ comes near the $P$, thus decreasing
$c$ such that in the heavy-quark limit, we get $c=0$. Thus our formula
found in Refs.\cite{RRS,OMRS} has now the {\it correct} Isgur-Wise limit.

Given that $c=0$ in the Isgur-Wise limit, we have the splitting
\be
\Delta E_{hf} = \frac{1}{2\I}  J_\ell ( J_\ell +1 ).
\label{energyp}
\ee
This $\Delta E_{hf}$ predicts
that there is an effective ``fine"
splitting of right sign and magnitude
between $\Lambda$ and the degenerate $\Sigma$ and $\Sigma^* $.
The predicted mass spectrum (denoting the mass by
the particle symbol) for b-quark baryons, with $\Lambda- N=4.65$ GeV
to fix $\alpha$, is
\be
{\Sigma_b} - N
= {\Sigma_b^*} - N &=& 4.84 \mbox{ GeV}.
\ee
These are comparable to the predictions of
quark potential models\,\cite{QM}
\be
({\Lambda_b} - N)^{\mbox{\tiny QM}}
&=& 4.65 \mbox{ GeV}, \nonumber \\
({\Sigma_b} - N)^{\mbox{\tiny QM}}
&=& 4.86 \mbox{ GeV}, \nonumber
\ee
and to those of bag models\,\cite{BM}
\be
({\Lambda_b} -  N)^{\mbox{\tiny BM}}
&=& 4.62 \mbox{ GeV}, \nonumber \\
({\Sigma_b} -  N)^{\mbox{\tiny BM}}
&=& 4.80 \mbox{ GeV}.\nonumber
\ee

It is possible, within the scheme described so far, to discuss hyperfine
splitting with a nonzero $c$. For a finite heavy-quark mass for which
$m_\Phi < m_\Phi^*$,
the CK model indicates that $c \sim 1/m_\Phi$. This is the hidden $m_\Phi^{-1}$
dependence buried in the hyperfine coefficient $c$ alluded above
which we conjecture may have an intricate connection to a Berry potential.
For a sufficiently large $m_\Phi$, we may therefore
assume $c=a/m_\Phi$. Now using
(\ref{hfsplit}), we can write for baryons with
one heavy quark $Q$\footnote{This splitting with $c_Q=0$
is equal to that of Ref.\cite{GLM}.
In Ref.\cite{GLM}, the authors predicted that $\Sigma_{\mbox{\tiny Q}}-
\Lambda_{\mbox{\tiny Q}}\approx \frac{2}{3} \Delta M = 195$ MeV
where $\Delta M$ is $M_\Delta - M_{\mbox{\tiny N}}$. They have no
$1-c_{\mbox{\tiny Q}}$ dependence in contrast to our eq. (\ref{finesplit}).}
\be
\Sigma_{\mbox{\tiny Q}}-\Lambda_{\mbox{\tiny Q}}=\frac{1}{\cal I}
(1-c_{\mbox{\tiny Q}})
\simeq 195 {\rm MeV} (1-c_{\mbox{\tiny Q}}). \label{finesplit}
\ee
With the experimental value $\Sigma_c-\Lambda_c\approx 168 {\rm MeV}$ for
the charmed baryons, we get $c_{c}\simeq 0.14$. This means that with
$m_{\mbox{\tiny D}}=1869 {\rm MeV}$, the constant $a\simeq 262$ MeV. So
\footnote{It is amusing to note that this formula works satisfactorily
even for the kaon for which one predicts $c_s \simeq 0.53$
to be compared with the empirical value 0.62.}
\be
c_{\Phi}\simeq 262 {\rm MeV}/m_\Phi.
\ee
Now for b-quark baryons, using $m_{\mbox{\tiny B}}=5279$ MeV, we find
$c_b\simeq 0.05$ which with (\ref{finesplit}) predicts
\be
\Sigma_b-\Lambda_b\approx 185 {\rm MeV}.
\ee
This agrees well with the quark-model prediction. Furthermore
the $\Sigma^*-\Sigma$
splitting comes out correctly also. For instance, it is predicted that
\be
\frac{\Sigma^*_b-\Sigma_b}{\Sigma^*_c-\Sigma_c}\simeq
\frac{m_{\mbox{\tiny D}}}{m_{\mbox{\tiny B}}}\approx 0.35
\ee
to be compared with the quark-model prediction $\sim 0.32$.
If one assumes that the heavy mesons $\Phi$ are weakly interacting, then
we can put more than one $\Phi$'s in the soliton and obtain the spectra
for $\Xi$'s and $\Omega$'s as reported in \cite{RRS,OMRS}. The agreement with
the quark-model results is surprisingly good as pointed out in those
references.

We have shown in this paper that one can interpolate the description
of baryon structure {\it smoothly} from light baryons (chiral symmetry)
to heavy baryons (Isgur-Wise symmetry) provided extra terms in chiral
Lagrangian are implemented to satisfy the Isgur-Wise symmetry. The results
obtained in Refs.\cite{JMW} and \cite{GLM} which build heavy-quark
skyrmions starting with a Lagrangian
that satisfies both chiral symmetry and Isgur-Wise
symmetry supplemented by symmetry-breaking terms of $O(1/m_Q)$
are strikingly similar to ours which start from chiral symmetry
with ``higher derivative" terms suitably added in to approach the heavy-quark
symmetry. The reason why the CK calculations
of heavy baryons of Refs.\cite{RRS,OMRS} with the principal contribution
coming from the pseudoscalar $P$ and minor contribution from the
vector $P^*$ (thus {\it apparently} possessing no manifest
heavy-quark symmetry) were successful may be that
the charm quark and bottom quark masses are not really heavy enough to require
heavy quark symmetry {\it ab initio}. This may be somewhat like the
strange quark mass which is heavy enough to be considered ``heavy"
in the sense of the Callan-Klebanov model
and light enough to be considered ``light"
in the sense of the Yabu-Ando model\,\cite{ya}. An attractive feature
of our results is that ours are interpretable in terms
of nonabelian Berry potentials.
\subsubsection*{Acknowledgments}
\indent

We acknowledge helpful correspondence on the subject with M.A. Nowak,
I. Zahed and N.N. Scoccola. This work is supported in part by the
Korea Science and Engineering Foundation through the CTP of Seoul
National University.

\end{document}